\documentclass[conference]{IEEEtran}

\newcommand{\pdftitle}{Characterizing and Modeling Distributed Training with Transient Cloud GPU Servers}

\newcommand{\para }[1]{\smallskip \noindent  {\bf #1}} 
\newcommand{\1}{{\em (i)}}
\newcommand{\2}{{\em (ii)}}
\newcommand{\3}{{\em (iii)}}
\newcommand{\4}{{\em (iv)}}

\usepackage{xspace}

\usepackage[table]{xcolor}
\usepackage[labelfont={bf,small},textfont={it,small}]{caption} 
\usepackage[labelformat=simple]{subcaption}

\usepackage[numbers,sort&compress]{natbib}
\usepackage{booktabs}
\usepackage{multirow}
\usepackage{amsmath}
\usepackage{balance}
\usepackage{graphicx}
\usepackage[hidelinks]{hyperref}  
\usepackage{float}
\usepackage{changes}

\newcommand{\sysname}{\textsc{CM-DARE}\xspace} 

\graphicspath{ {figures/} }

\newcommand{\eat}[1]{}
\newcommand{\done}[1]{}

\newcommand*\ttvar[1]{\texttt{\expandafter\dottvar\detokenize{#1}\relax}}
\newcommand*\dottvar[1]{\ifx\relax#1\else
  \expandafter\ifx\string_#1\string_\allowbreak\else#1\fi
  \expandafter\dottvar\fi}

\renewcommand{\paragraph}{\para}
\let\labelindent\relax
\usepackage{enumitem}

\begin{document}

\bstctlcite{IEEEexample:BSTcontrol}

\title{\pdftitle}

\author{
\IEEEauthorblockN{Shijian Li}
\IEEEauthorblockA{Worcester Polytechnic Institute, USA\\
$sli8@wpi.edu$}
\and
\IEEEauthorblockN{Robert J. Walls}
\IEEEauthorblockA{Worcester Polytechnic Institute, USA\\
$rjwalls@wpi.edu$}
\and
\IEEEauthorblockN{Tian Guo}
\IEEEauthorblockA{Worcester Polytechnic Institute, USA\\
$tian@wpi.edu$}
}

\maketitle

\thispagestyle{plain}
\pagestyle{plain}

\begin{abstract}
Cloud GPU servers have become the \emph{de facto} way for deep learning
practitioners to train complex models on large-scale datasets.
However, it is challenging to determine the appropriate cluster
configuration---e.g., server type and number---for different training workloads while balancing the trade-offs in training time, cost, and model accuracy. 
Adding to the complexity is the potential to reduce the monetary cost by using
cheaper, but revocable, transient GPU servers. 

In this work, we analyze distributed training performance under diverse cluster configurations using \sysname, a cloud-based measurement and training framework. 
Our empirical datasets include measurements from three GPU types, six geographic regions, twenty convolutional neural networks, and thousands of Google Cloud servers.
We also demonstrate the feasibility of predicting training speed and overhead using regression-based models. 
Finally, we discuss potential use cases of our performance modeling such as detecting and mitigating performance bottlenecks.
\end{abstract}


\section{Introduction}
\label{sec:intro}

The process of training deep neural networks (DNNs) has evolved from using
single-GPU servers~\cite{shi2018performance} to
distributed GPU clusters~\cite{firecaffe,geeps} that can support larger and more
complex DNNs.
Cloud computing, providing on-demand access to these critical yet expensive GPU
resources, has become a popular option for practitioners. 
Today's cloud provides its customers abundant options to configure the
training clusters, presenting opportunities for tailoring resource
acquisition to the specific training workload. 
When using 
cloud-based GPU servers to train deep learning models, one can choose the
server's CPU and memory, specify the GPU type, decide the number of servers, as
well as pick the desired datacenter location. 
However, this configuration flexibility also imposes additional complexity upon deep learning practitioners. 

Concurrently, to lower the
monetary cost of training, one could also consider using a special type of cloud
servers, referred to as \emph{transient servers}, that have lower unit costs 
with the caveat that the server can be revoked at any time~\cite{ec2_spot,gce_preemptible}.
Revoked GPU servers often mean
significant loss of work and require manual effort by the practitioner to request
new servers, to reconfigure the training cluster, and even to diagnose potential
performance bottlenecks. Concretely, when a GPU server is revoked, all its
local training progress will disappear and in the worst case, the revocation will
also impede the functionality of saving the trained model~\cite{tensorflow,2019icac:speedup}. 

In this work, we set out to characterize and predict the impact of cluster
configuration on distributed training, in the context of transient and
traditional \emph{on-demand} cloud servers. We measured and characterized several key factors that
impact distributed training on
transient servers and evaluated regression-based models for
predicting training throughput and fault-tolerance overhead. 

To streamline measurement and data collection on distributed training, we
designed and built a framework called \sysname. It allows us to measure, monitor,
and collect metrics such as training speed and revocation time, which supports
our performance characterization and modeling and enables use cases such as
performance bottleneck detection.
We built \sysname on top of an existing distributed training framework
(\emph{TensorFlow}~\cite{tensorflow}) and library
(\emph{Tensor2Tensor}~\cite{tensor2tensor}), with transient-specific optimizations that
mitigate the impact of revocation and improve fault-tolerance. Though we
exclusively used \emph{TensorFlow} and Google Cloud in
this work, we argue 
that our measurement methodology (e.g., the use of custom
convolutional neural networks) can be extended to other deep learning frameworks
and cloud providers.

Our work differs from prior work in distributed training performance
modeling in three key aspects. \emph{First,} it consists of large-scale,
cloud-based measurement and data-driven performance modeling rather than
theoretical modeling and on-premise
measurement~\cite{dl_perf1,qi:iclr17:paleo,shi2018performance}. 
\emph{Second,} we identified use cases that benefit from having access to the
raw measurement data, performance models, and \sysname measurement infrastructure.
\emph{Finally,} we are the first to characterize and model 
performance of distributed training with \emph{transient servers}. In short, we make the following contributions.

\begin{itemize}[leftmargin=.12in,topsep=4pt]
    \setlength\itemsep{0.1em}
    \item We conducted a large-scale measurement study that includes twenty
    convolutional neural networks on three types of Google Cloud GPU servers. We observe, for example, that the training speed of heterogenous clusters---i.e.,
    clusters consisting of different GPU hardware---is approximately the sum of
    individual server speeds. 
    Our dataset and \sysname are available in the project
    GitHub repository\footnote{\url{https://github.com/cake-lab/CM-DARE}}. 
    \item We built and evaluated performance models that predict the
    training speed and fault-tolerance overhead of GPU clusters with
    as low as 3.4\% mean absolute percentage error. Such models serve as the
    building blocks for predicting heterogeneous cluster training performance.
    More importantly, we identified appropriate deployment scenarios for each performance model.
    \item We identified use cases, such as detecting and mitigating distributed
    training performance bottlenecks, that would benefit from our prediction
    models. 
    \item We designed and implemented a measurement and training
    framework called \sysname, which simplifies distributed training on
    transient servers and improves the robustness of existing fault-tolerance
    mechanisms.
    \end{itemize}

\section{Overview of the \sysname Framework}
\label{sec:Overview}

\begin{figure}[t]
    \centering
    \includegraphics[width=0.48\textwidth]{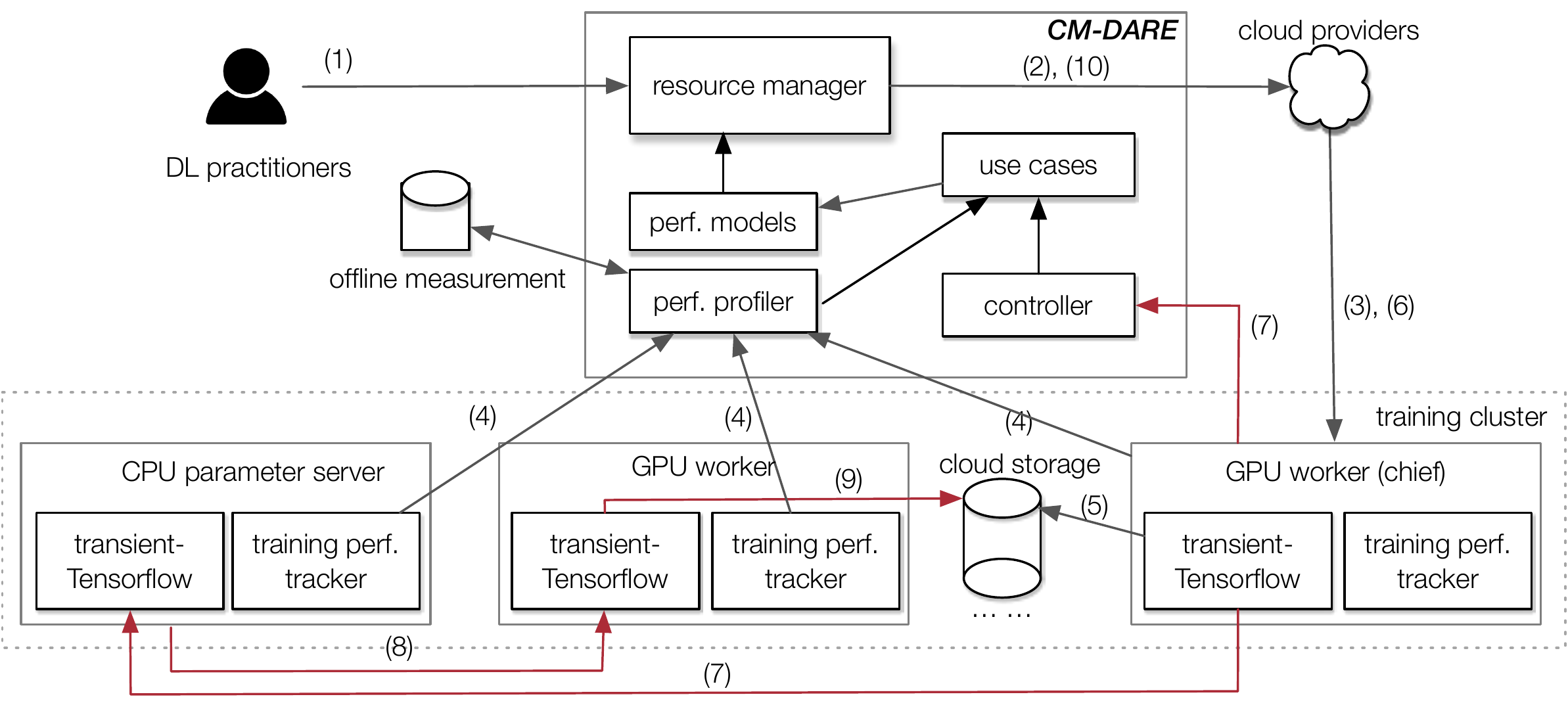}
    \caption{\textbf{\sysname architecture and workflow.} 
    The key workflow for performing transient-aware distributed training is
    labeled numerically, with the revocation-triggered interactions highlighted
    in red.
    }
    \label{fig:spottrain_arch}
\end{figure}


\sysname is a measurement framework we built and used to characterize and
predict the performance of training convolutional neural
networks (CNNs) on clusters of cloud-based GPU servers, i.e., \emph{distributed
training}. 

Specifically, we focus on \emph{asynchronous
training with parameter servers}, a popular distributed training architecture
implemented by Google's TensorFlow~\cite{tensorflow} and 
commonly used for models that can fit into the memory of a discrete GPU.
In this architecture, servers are separated logically
into two categories: \emph{parameter servers} and \emph{workers}. Parameter
servers update the deep learning model parameters after each
GPU server (i.e., worker) generates the gradients. 
Each worker holds its own copy of the
entire deep learning model and works on subsets of the training dataset.
The training is asynchronous
because each worker communicates with the parameter servers at its own pace.
One worker is designated as the \emph{chief worker} and is given additional responsibilities, including
periodically saving model parameters to cloud storage, i.e., \emph{checkpointing}.

The asynchronous nature of this architecture offers two key benefits for transient distributed training.
First, it is resilient to transient revocations because the cluster can continue
training even if a worker is revoked. 
Second, it reduces the impact of hardware differences in heterogeneous clusters
because slower workers do not impede others. 

At the core of \sysname, depicted in Figure~\ref{fig:spottrain_arch}, is the \emph{transient-aware performance models} which
is powered by \emph{performance profiler} that continuously monitors training
performance and transient server revocations. In addition, \sysname
includes \emph{transient-TensorFlow}, a modified version of TensorFlow,
that handles worker revocations by notifying the parameter server and supporting
checkpointing even when the chief worker is revoked.

To collect measurements with \sysname, (1) we provide a training script with
information such as cluster configuration, which (2) the \emph{resource manager}
uses for setting up the cloud training cluster. (3) All training servers,
including on-demand parameter servers and transient GPU workers, will
run \emph{transient-TensorFlow} which establishes RPC connections between
parameter servers and workers and (4) the \emph{performance tracker} that sends
training performance to the \emph{performance profiler}. (5) After the specified
checkpoint interval, the chief worker saves the current model parameters to
cloud storage. 
(6) In the case that the chief worker is revoked,
(7) the chief will notify the parameter server, as well as the \emph{controller}, about its revocation. 
(8) The parameter server will then select one GPU worker to take
over checkpointing, (9) and the worker will save the checkpoint to the same
cloud storage at the specified interval. (10) The \emph{resource manager}
fulfills cluster configuration changes that are determined by the 
\emph{controller} based on the specified use cases, performance
models, and online measurement. Finally, we obtain the trained
models and measurement data once the training is completed.

Currently, \sysname runs on Google Cloud.
We chose Google Cloud because it allows customization of GPU
servers, which provides better control and flexibility for training deep
learning models with different resource requirements. 
Further, Google's transient servers, called preemptible VMs in the Google Cloud
argot, have a maximum lifetime of 24 hours and are offered at fixed 
prices that are significantly lower than their on-demand counterparts. 

We characterize and predict distributed training performance in the context of
training speed in Section~\ref{sec:training_speed}, fault-tolerance
overhead in Section~\ref{subsec:fault_tolerance_overhead}, and revocation
overhead in Section~\ref{sub:revocation_overhead}. Finally, in
Section~\ref{sec:use_cases}, we explore two potential use cases that could
benefit from our study: predicting
training speed of heterogeneous clusters and detecting training bottlenecks.

\section{Understanding and Predicting Training speed}
\label{sec:training_speed}

Understanding how training speed varies based on key factors such as \emph{GPU
server type} and \emph{model characteristics}, is the first
step toward predicting distributed training performance. In this section, we
quantify such relationships with \sysname-enabled empirical measurements. 
In summary, we find that regression-based prediction is a promising approach due to the strong
correlation between training speed, GPU computational capacity, and model
complexity. Further, the limited selection of available cloud GPUs make it
feasible to build predictive models for individual GPU types and thus achieve
higher prediction accuracy. Moreover, the training speed of an entire cluster is
approximately the sum of individual worker speeds until a
parameter-server-based bottleneck is reached. 
Finally, compared to prior approaches that do not consider transient server
revocations and assume stable training environment~\cite{peng2018optimus, lin2018model, zheng2019cynthia}, our data-driven approach
achieves low prediction error of 9\%. 


\begin{table}[t]
    \sf
    \centering
    \caption{\textbf{Training speed (in steps per second) for the simplest cluster configuration.} We measured a cluster consists of one GPU worker and one parameter server in the same data center. The GFLOPs of CNN models are calculated based on \emph{CIFAR-10} dataset.}
    \label{tab:training_speed_gpu}
    \resizebox{0.45\textwidth}{!}{
    \begin{tabular}{@{}r rrrr@{}}
        \toprule
        \multicolumn{1}{c}{}                                              & \multicolumn{4}{c}{\textbf{CNN model (GFLOPs)}}                         \\ \cmidrule(l){2-5}
        \textbf{\begin{tabular}[c]{@{}r@{}}GPU\\ (teraflops)\end{tabular}} & \textbf{\begin{tabular}[c]{@{}r@{}}ResNet-15 \\ (0.59)\end{tabular}} & \textbf{\begin{tabular}[c]{@{}r@{}}ResNet-32\\ (1.54)\end{tabular}} & \textbf{\begin{tabular}[c]{@{}r@{}}Shake Shake small\\ (2.41)\end{tabular}} & \textbf{\begin{tabular}[c]{@{}r@{}}Shake Shake Large\\ (21.3)\end{tabular}} \\ \midrule
        \textbf{K80 (4.11)}                                                & 9.46 $\pm$ 0.19                                                      & 4.56 $\pm$ 0.08                                                     & 2.58 $\pm$ 0.02                                                             & 0.70 $\pm$ 0.002                                                            \\
        \textbf{P100 (9.53)}                                               & 21.16 $\pm$ 0.47                                                     & 12.19 $\pm$ 0.41                                                    & 6.99 $\pm$ 0.35                                                             & 1.98 $\pm$ 0.03                                                             \\
        \textbf{V100 (14.13)}                                              & 27.38 $\pm$ 0.88                                                     & 15.61 $\pm$ 0.38                                                    & 8.80 $\pm$ 0.24                                                             & 2.18 $\pm$ 0.04                                                             \\ \bottomrule
        \end{tabular}
    }
    \end{table}


\subsection{Measurement Methodology}
\label{subsec:trainingspeed:methodology}

\paragraph{Training Dataset.}
We chose \emph{CIFAR-10}, one of the most widely used datasets in deep learning
research, as the training dataset~\cite{cifar10}. \emph{CIFAR-10} contains a
total of 60K images with dimensions of 32 $\times$ 32 pixels. The training
workload is provided by practitioners in the form of \emph{number of steps} which each step
goes through a mini-batch of images. 
Larger-scale datasets, such as \emph{ImageNet}, that are commonly used for
improving the real-world model accuracy, were unnecessary as our
measurements focus on training speed.

\paragraph{Models.}
We used two \emph{ResNet}~\cite{resnet} and two \emph{Shake
Shake}~\cite{gastaldi2017shake} implementations from the
\emph{Tensor2Tensor} framework. 
These four CNN models are popular for image
classification and have different characteristics such as model complexity
that are useful for our study. 
\emph{Model complexity} is defined as the number of
floating point operations (FLOPs) required by the CNN model to train on one image. 
We further generated an additional 16 variants of CNN models by varying the
number of hidden layers and the size of each hidden layer; these custom models
allowed us to better observe how model complexity impacts training time. 
We used the built-in TensorFlow profiler
tool to calculate the FLOPs for each model.

\paragraph{GPU Types.} We used three GPU types offered by Google Cloud: Nvidia
Tesla \emph{K80}, \emph{P100}, and \emph{V100}. 
These GPUs used PCIe and had 12GB, 16GB, and 16GB of memory, respectively. 
They had computational capacity of 4.11, 9.53, and 14.13 teraflops. We
chose these GPU types because they are the only three offered by Google Cloud
that are commonly used for training. We refer to a server with access to a GPU as a \emph{GPU server}.
Each GPU server was configured with 4 vCPUs and 52GB of main memory.
During our experiments, neither the CPU nor the main memory were saturated.

\paragraph{Cluster Configuration.} For measuring the impact of model and GPU type, we used a
simple cluster consisting of one GPU server and one parameter server with both
servers residing in the same data center. 
We ran the parameter server on a non-revocable server with 4vCPUs, 16GB
of main memory, and Ubuntu 18 LTS. GPUs were not needed for the parameter
server as its primary tasks, aggregating gradients and updating parameters, are
less computation-intensive and are often bound by network communication~\cite{jeffdean}.
We also evaluated different cluster configurations by mixing the type of GPU
servers with varying number.

\paragraph{Measuring Training Speed.} We utilized built-in TensorFlow
functionality to log 
training speed of the entire cluster. Training speed is defined as \emph{steps per second} where each
step involves the generation of gradients based on the new model parameters
using a batch of images. Unless otherwise specified, we averaged the training
speed every 100 steps.
For each cluster, we trained and recorded for 4000 steps. We used the same
training workload for all clusters and set
the checkpoint interval to be larger than our measurement
duration to avoid measuring checkpoint overhead, which we consider in
Section~\ref{subsec:fault_tolerance_overhead}.
To measure the training speed of individual workers, without incurring logging
overhead associated with hook
functions, we used the TensorFlow \emph{TFProf} tool. 


\begin{figure}[t]
    \begin{center}
        \includegraphics[width=0.4\textwidth]{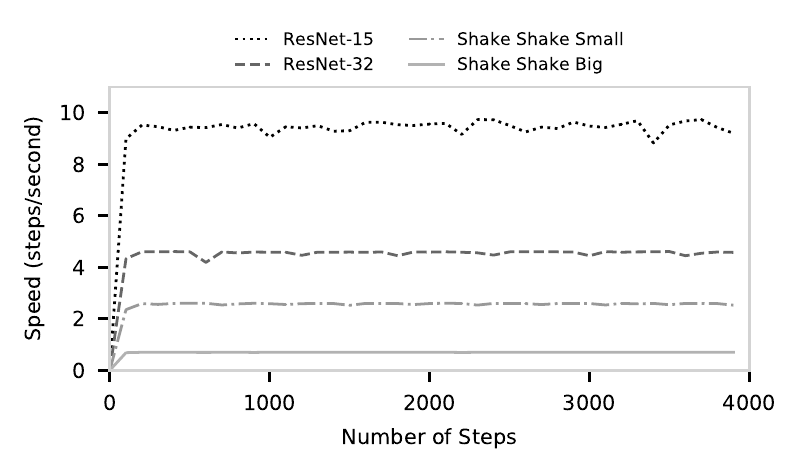}
    \end{center}
    \caption{
\textbf{Training speed for the simplest cluster configuration (K80).}
We plotted the training
speed of all four representative CNN models, and observed that
training speed is rather stable after warmup.}
    \label{fig:steady_speed}
\end{figure}

\subsection{Impact of Model and GPU Type} 
\label{subsec:trainingspeed:gpu_and_cnn}

Table~\ref{tab:training_speed_gpu} shows the average training speed (and
standard deviation) for different combinations of model and GPU type. To avoid
including noisy data, we discarded the
measurements associated with the \emph{first 100} steps. As expected, the higher
the computational capacity of the GPU, the faster the training speed.
For example, the \emph{V100} server has the highest training speed for all
four CNN models. Further, the training speed drops as model
complexity increases. For instance, training a \emph{ResNet-32} of 1.54 GFLOPs
is almost 2X slower than training a \emph{ResNet-15} of 0.59 GLOPs using
the same \emph{K80} GPU server.

Another important observation, visualized in Figure~\ref{fig:steady_speed} for
a \emph{K80} server, is that training speed was stable after the warm-up
period, with a maximum coefficient of variation of 0.02.
We observed similar behavior for the other two types of GPU servers.
This training speed consistency has several important implications, namely the
feasibility of predicting the speed using historical data and the possibility
to quickly detect (and address) under-performing workers.


\begin{figure}[t]
    \captionsetup[subfigure]{aboveskip=-1pt,belowskip=-1pt}
    \begin{subfigure}{0.23\textwidth}
    \centering
    \includegraphics[width=\textwidth]{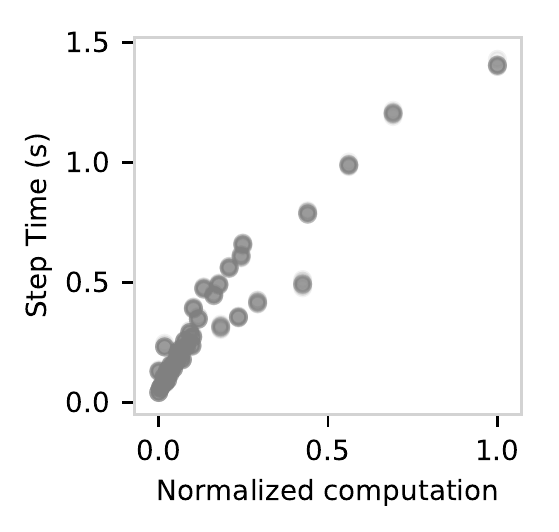}
    \caption{Normalized computation}
    \label{subfig:step_time_comp_1}
    \end{subfigure}
    \hfill
\begin{subfigure}{0.235\textwidth}
    \centering
    \includegraphics[width=\textwidth]{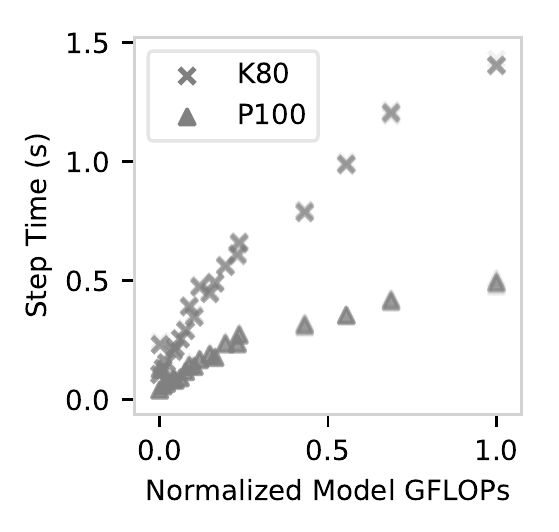}
    \caption{Normalized model complexity}
    \label{subfig:step_time_comp_2}
    \end{subfigure}
    \caption{\textbf{Step time vs. normalized computation ratio $C_{norm}$ and model complexity $C_m$.} 
    We observed strong positive correlation between average \emph{step time} and
    normalized computation and model complexity for both \emph{K80} and
    \emph{P100} GPU servers.
    }
    \label{fig:step_time_comp}
\end{figure}


\subsubsection{Predicting the Impact}
\label{subsubsec:trainingspeed:predicting}
The next question we explore is how to leverage the above observations to
predict the training speed of an individual worker, especially when training a
previously unobserved CNN model. In Section~\ref{subsec:cluster_size}, we
further investigate the question of how to predict the training speed of an
entire cluster.     

Figures~\ref{subfig:step_time_comp_1} and \ref{subfig:step_time_comp_2} show the
relationship between \emph{step time} $S$ and 
normalized computation ratio $C_{norm}$ and normalized model complexity $C_m$,
respectively. 
The \emph{computation ratio} is defined as model complexity divided by GPU
computational capacity $C_{gpu}$, and \emph{step time} is the inverse of
training speed.
The computation ratio and model complexity were normalized using min-max normalization.\footnote{
We also considered \emph{z-score standardization} for preprocessing;
however as our data does not follow a Gaussian distribution, it would be less
beneficial to apply this technique.} 
Each dot represents the observed step time,
averaged over 1400 steps, from training a CNN model. We collected data for a set
of twenty CNN models, comprising the 4 models used for the
observations in the previous section and the 16 custom models mentioned in the methodology.

We make two key observations. 
\emph{First,} the step time of different GPUs form a trend line when
using $C_{norm}$ and are distinctly separated when
using $C_m$. This suggests that 
both normalized computation ratio and normalized model complexity are useful in
predicting training speed.
Further, it implies the benefit for building performance models for different
GPU types.
\emph{Second,} the shapes of the trend lines indicate that linear functions
might be fitted for predicting the step time. 


\begin{table}[t]
    \centering
    \caption{
    \textbf{Comparison of step time prediction models.}
    We evaluated the listed regression models in predicting \emph{step time},
    using k-fold cross validation MAE and test dataset MAE. The
    Unit for MAE is seconds.}
\label{tab:speed_pred}
    \resizebox{0.45\textwidth}{!}{
    \begin{tabular}{@{}lrrr@{}}
        \toprule
        \textbf{Regression Model}            & \multicolumn{1}{l}{\textbf{Input Feature}} & \multicolumn{1}{l}{\textbf{K-fold MAE}} & \multicolumn{1}{l}{\textbf{Test MAE}} \\ \midrule
        \textbf{Univariate, GPU-agnostic}                  & $C_{norm}$                                       & 0.072 $\pm$ 0.015                       & 0.068                                 \\
        \textbf{Multivariate, GPU-agnostic}                & $C_{m}$, $C_{gpu}$                         & 0.103 $\pm$ 0.026                       & 0.093                                 \\
        \textbf{Univariate, K80}            & $C_{m}$                                    & 0.065 $\pm$ 0.013                       & 0.068                                 \\
        \textbf{SVR Polynomial Kernel, K80} & $C_{m}$                                    & 0.035 $\pm$ 0.014                       & 0.041                                 \\
        \textbf{SVR RBF Kernel, K80}        & $C_{m}$                                    & 0.026 $\pm$ 0.012                       & 0.031                                 \\ 
        \textbf{Univariate, P100}            & $C_{m}$                                    & 0.029 $\pm$ 0.008                       & 0.031                                 \\
        \textbf{SVR Polynomial Kernel, P100} & $C_{m}$                                    & 0.019 $\pm$ 0.007                       & 0.020                                 \\
        \textbf{SVR RBF Kernel, P100}        & $C_{m}$                                    & 0.012 $\pm$ 0.008                       & 0.016                                 \\ 
        \bottomrule
        \end{tabular}}
\end{table}

Based on the observations above, we evaluated eight regression models, listed
in Table~\ref{tab:speed_pred}, for predicting training speed. 
We chose a mix of univariate, multivariate, and support vector
regression (SVR) models because
the former two are simple and commonly used, and the latter has been shown to work well in modeling performance in cloud environments~\cite{kundu2012modeling}.
These models can be divided into two categories: GPU-agnostic and GPU-specific.
The GPU-agnostic univariate regression is modeled as $S = a
\times C_{norm} + b$, while the GPU-agnostic multivariate regression is modeled
as $S = a \times C_{m} + b \times C_{gpu} + c$, where $a, b, c$ are learned
parameters.

Training GPU-specific prediction models are feasible because 
cloud GPUs are often limited in selections and are usually
not customizable. Specifically, we considered the following three GPU-specific 
regression models: 

\setlength{\abovedisplayskip}{-5pt}
\setlength{\belowdisplayskip}{1pt}
\begin{align} 
    S_{gpu} &= a \cdot C_{m} + b, \label{eq:trainingspeed:univariate}\\ 
    S_{gpu} &= \sum^N_{i=1}(\alpha_{i} - \alpha_{i}^{*}) \cdot {(C_{m_{i}}, C_{m})}^2, \label{eq:trainingspeed:svrpoly}\\
    S_{gpu} &= \sum^N_{i=1}(\alpha_{i} - \alpha_{i}^{*}) \cdot \exp(- \frac{{ || C_{m_{i}} - C_{m}} ||^2}{2 \sigma^2}), \label{eq:trainingspeed:RBF} 
\end{align}

where $S_{gpu}$ denotes the \emph{step time} of one specific GPU;
$\alpha_{i}$ and $\alpha_{i}^{*}$ are Lagrange multipliers used in SVR to
determine support vectors; and ${(C_{m_{i}}, C_{m})}^2$ and $\exp(- \frac{{ ||
C_{m_{i}} - C_{m}} ||^2}{2 \sigma^2})$ are two-degree
polynomial and RBF kernel functions, respectively.

For training each regression model, we randomly split the dataset into 
training data and test data with 4:1 ratio. 
We conducted k-fold cross validation on the training data, and evaluated the
performance of resulting regression models using mean absolute error (MAE) for
both training and test data.  
We chose MAE because it provides a more
natural and unambiguous measurement compared to other metrics such as root mean
square error (RMSE)~\cite{willmott2005advantages}. 
Further, \emph{k-fold MAE} allows us to compare against different regression models 
and \emph{test MAE} provides insight regarding the robustness of each regression model.
For training SVR-based models, we used grid search cross validation 
to search for the optimal set of hyperparameters, i.e., penalty $p$ and $\epsilon$, that yield the best MAE. We followed common practice and set the range for $p$ to be $[10, 100]$ with a step increment of 10, and $\epsilon$ to be $[0.01,  0.1]$ with a step increment of 0.01.

As shown in Table~\ref{tab:speed_pred}, the GPU-specific regression models
achieved higher MAE than the GPU-agnostic predictive models. For
example, all six GPU-specific models had a MAE of less than
$0.068$ seconds on the test dataset. As the average \emph{step time} across
different CNN models is $0.48$ seconds, we believe 
models with such MAEs can produce reasonable predictions. In comparison, the
GPU-agnostic regression models had up to $0.093$ seconds MAE on the test dataset. 
Furthermore, the SVR models with the non-linear RBF kernel function provided
a better fit than those with the polynomial kernel function---providing, for
example, the best MAE for both k-fold cross validation and test dataset.
The mean absolute percentage error (MAPE) on test dataset for the K80-specific SVR model with RBF kernel was 9.02\%, compared to 13.79\% for the P100-specific SVR model with polynomial kernel.

\subsection{Impact of Cluster Size and Heterogeneity on Worker Speed}

\begin{table}[t]
    \centering
    \sf
    \caption{\textbf{Average step time (in millisecond) of an individual worker when training ResNet-32.} 
    We studied the impact of adding more GPU servers on a single GPU's training
    speed. Clusters are represented as $(x, y, z)$ where x, y, z denote the
    number of \emph{K80}, \emph{P100}, \emph{V100} GPU servers respectively.
    }
    \label{tab:cluster_training_speed}
    
    \resizebox{0.45\textwidth}{!}{
    \begin{tabular}{@{}llllll@{}}
        \toprule
        \textbf{}     & \multicolumn{1}{c}{\textbf{Baseline}}  & \multicolumn{3}{c}{\textbf{Homogeneous}}                                                                                 & \textbf{Heterogeneous}                 \\ \cmidrule(l){2-2}\cmidrule(l){3-5}\cmidrule(l){6-6}
        \textbf{}     & \multicolumn{1}{r}{\textbf{(1, 0, 0)}} &
        \multicolumn{1}{r}{\textbf{(2, 0, 0)}} & \multicolumn{1}{r}{\textbf{(4,
        0, 0)}} & \multicolumn{1}{r}{\textbf{(8, 0, 0)}} &
        \multicolumn{1}{r}{\textbf{(2, 1, 1)}}   \vspace{.05in}        
        \\ \midrule
        \textbf{K80}  & 229.85 $\pm$ 3.04                      & 232.08 $\pm$ 2.22                      & 229.57 $\pm$ 3.15                      & 227.46 $\pm$ 5.06                      & 221.16 $\pm$ 2.66                      \\
        \textbf{P100} & 105.45 $\pm$ 1.99                      & 105.27 $\pm$ 1.45                      & 112. 73 $\pm$ 6.52                     & 198.11 $\pm$ 18.65                     & 107.61 $\pm$ 2.13                      \\
        \textbf{V100} & 92.38 $\pm$ 3.64                       & 95.90 $\pm$ 4.07                       & 106.36 $\pm$ 6.16                      & 191.72 $\pm$ 26.38                     & 93.52 $\pm$ 4.58                       \\ \bottomrule
        \end{tabular}
    }
    \end{table}

To predict training speed for an entire
cluster, we must first understand the impact of cluster size and mixing GPU
types on the training speed of an individual worker. 
Table~\ref{tab:cluster_training_speed} shows the average step
time for individual \emph{K80}, \emph{P100}, and \emph{V100} workers when used
as part of both homogeneous and heterogeneous clusters. 
The baseline column shows the average step time for a cluster consisting of a
single worker.

We make three key observations. 
\emph{First,} for homogeneous clusters, the average training speed of an
individual worker was roughly the same until the cluster became large enough to
encounter a parameter server bottleneck.
This bottleneck arises when the rate of workers' output (i.e., computed
gradients) exceeds the parameter server's capacity. Consequently, the training
is bounded
by how fast the parameter server can update model parameters. 
Notice that the \emph{K80} workers, with the least powerful GPU, did not reach
this bottleneck in our experiments and the average
step time was within 1\% for all tested cluster sizes. 
In contrast, workers with the more powerful GPUs hit this bottleneck at smaller
cluster sizes (8 for \emph{P100} and 4 for \emph{V100}). 
We discuss how to mitigate the impact of parameter
server bottlenecks in Section~\ref{sec:uescase:bottleneck}.

\emph{Second,} as the cluster size increases,
we observe higher variations for the average step time. For example, the coefficient of
variation increases from $0.019$ to $0.094$ for \emph{P100} clusters. 
\emph{Third,} the use of heterogeneous clusters does not appear to impact the
training speed of an individual worker. For instance, the average step time of a
\emph{V100} worker is 92.38ms in the baseline
cluster and 93.52ms in the heterogeneous cluster.


\begin{figure}[t]
    \begin{center}
        \includegraphics[width=0.4\textwidth]{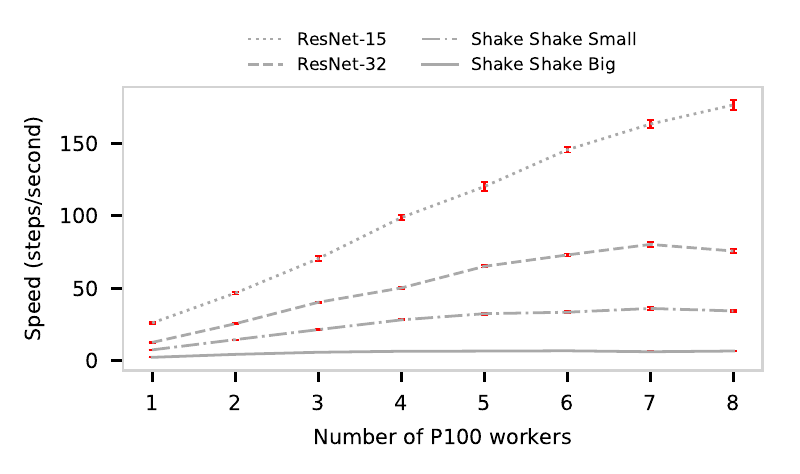}
    \end{center}
    \caption{
    \textbf{Empirically measured cluster training speed.}
    We observed that training speed bottleneck can be reached faster for more
    complex CNN models and powerful GPUs such as \emph{P100}.
    \emph{Shake Shake} models exhibit negligible variations
    compared to \emph{ResNet} models, potentially due to higher model
    computation complexity saturating GPU.}
    \label{fig:trainingspeed:clustersize}
\end{figure}

\subsection{Impact of Cluster Size on Cluster Training Speed}
\label{subsec:cluster_size}

To understand the impact of the number of GPU servers on the cluster training
speed, we trained the four Tensor2Tensor models with clusters comprised of an 
increasing number of \emph{P100} GPU servers.
Figure~\ref{fig:trainingspeed:clustersize} shows the average training speed for
each cluster. 

We make three key observations.
\emph{First,} the cluster training speed increases as the cluster size grows.
The upward trend is most obvious for \emph{ResNet-15}, the least
computationally-intensive model of the four. 
\emph{Second,} for both \emph{ResNet-32} and \emph{Shake Shake Small} models,
the training speed starts to plateau after more than four GPU servers
in the cluster, caused by the parameter server bottleneck discussed previously. 
\emph{Third}, the lack of training speed improvement for \emph{Shake Shake Big},
the most complex of the four models, suggests that the computational capacity of
the \emph{P100} GPU was insufficient for the model.
In a separate experiment, not shown, we observed a positive correlation between
the training speed and cluster size 
for \emph{Shake Shake Big} after switching from \emph{P100} to the more powerful
\emph{V100} GPU. 

All the observations in this section 
indicate that we can effectively predict unknown clusters'
training speed by leveraging our understanding of individual worker's
performance, and composing from our previously built performance models.
Further, if the predicted performance deviates from the online measurement,
\sysname can flag parameter servers as the bottleneck and start provisioning
additional parameter servers. 
\section{Modeling Fault-tolerance Overhead}
\label{subsec:fault_tolerance_overhead}

Current deep learning frameworks, such as TensorFlow, often provide basic
fault-tolerance mechanisms. For example, TensorFlow allows deep learning
practitioners to periodically save the most recent model parameters to remote
storage. These model files serve as an intermediate result, and allow resuming
the training from the checkpoint in case of a failed
training session. 
Fault-tolerance mechanisms are especially important when using transient servers
for distributed training, as these mechanisms can reduce the amount of work
loss when a worker is revoked. 

In this section, we study how fault-tolerance mechanisms, specifically
checkpointing CNN models, impact distributed training time. Our observations
indicate that the tasks of training and checkpointing happen sequentially and
that one can take into account the checkpoint overhead by directly adding to the
predicted training time. Our checkpoint prediction models
yield only 5.38\% mean absolute percentage error and our
analysis suggests the value of using different prediction models in different
deployment scenarios. 

\subsection{Measurement Methodology}
\label{sec:checkpoint:methodology}

\paragraph{Checkpoint-related Files.} 
TensorFlow generates three types of files, i.e., \emph{data, index and
meta} files, when checkpointing deep learning models. Both index file and
meta file sizes are highly correlated to the number of tensors, e.g., vectors or
matrices, in the CNN
model. We denote the size of data, meta, and index files with $S_d$, $S_m$, and
$S_i$, respectively, and use $S_c$ to denote the sum of these three files.

\paragraph{Measuring Checkpoint Time.} We instrumented the checkpointing function used
by TensorFlow and measured the time to checkpoint all twenty CNN
models described in Section~\ref{subsec:trainingspeed:methodology}.
In TensorFlow, the chief worker is responsible for checkpointing for the entire
cluster. Further, checkpointing does not run on the GPU. Consequently, we
measured the checkpointing time 
using a cluster consisting of a parameter server and a single \emph{K80}
worker, i.e., the chief worker. 
To minimize the network impact on the measured checkpointing time, we configured
the worker to save checkpoints to remote storage in the same data center
as the training cluster. 

\subsection{Understanding Checkpoint Time}
\label{sec:checkpoint:understanding}


\begin{figure}[t]
    \centering
    \includegraphics[width=0.4\textwidth]{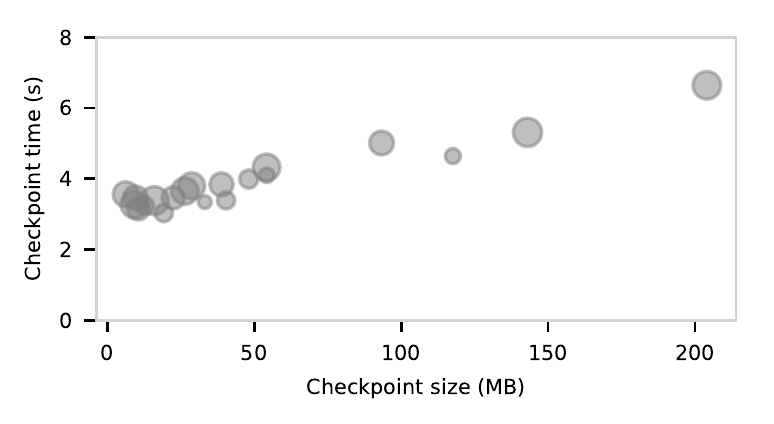}
    \caption{\textbf{Checkpoint duration vs. checkpoint size.} We measured the
    total checkpoint time for all twenty CNN models for five times and plotted
    both the average and coefficient of variation (circle size).}
    \label{fig:ckpt_dura}
\end{figure}

Figure~\ref{fig:ckpt_dura} shows the checkpoint time, averaged over five
checkpoints, for all twenty CNN models. We observed a low coefficient of
variation for all models, ranging from 0.018 to 0.073, and a positive
correlation between checkpoint size and time. 
By cross-examining the training speed with and without
checkpointing, we confirmed that the tasks of checkpointing and training
are conducted in sequence. 
As an example, in the case of training \emph{ResNet-32}, it takes $25.64$ versus $21.93$ seconds on average to finish
100 steps with and without checkpointing, respectively. The difference of
$3.71$ seconds is consistent with the measured \emph{ResNet-32} checkpoint
time of $3.84\pm0.25$ seconds. 
This indicates that we can directly add the checkpoint
overhead to the distributed training time modeled without checkpointing. 
Finally, recall that only \emph{one} worker performs the checkpointing. As such,
we just need to account for one interrupted worker when predicting the overhead. 

\subsection{Predicting Checkpoint Time}
\label{sec:checkpoint:prediction}

We considered the four regression models listed in Table~\ref{tab:ckpt_pred} for
predicting checkpoint time. 
Further, given that
\emph{index} and \emph{meta} file sizes are both correlated to the number of
tensors, we use principal component analysis (PCA) to preprocess the input
features to automatically reduce the variable dimensions to two components.
Similar to predicting training speed
(Section~\ref{subsec:trainingspeed:gpu_and_cnn}), we considered the following
models: \1 $T_{c} =
a \cdot S_c + b$, \2 $T_{c} = a \cdot S_d + b \cdot S_m + c$, \3 $T_c = (a, b)
\cdot PCA(S_d, S_m, S_i) + c$, \4 $T_c= \sum^N_{i=1}(\alpha_{i} -
\alpha_{i}^{*}) \cdot \exp(- \frac{{ || S_c^i - S_c} ||^2}{2 \sigma^2})$,
where $T_c$ denotes checkpointing time; $\alpha_{i}$ and $\alpha_{i}^{*}$ are
Lagrange multipliers used in SVR to determine support vectors; and
$\exp(-\frac{{ || S_c^i - S_c} ||^2}{2 \sigma^2})$ is RBF kernel function, respectively.


\begin{table}[t]
    \centering
    \sf 
    \caption{\textbf{Comparison of checkpoint prediction models.} 
    We evaluated the listed regression models in predicting checkpointing time,
    using k-fold cross validation MAE and test dataset MAE. $S_c$, $S_d$, $S_m$,
    and $S_i$ denote
    the file sizes for total checkpoint, data, meta, and index
    files, respectively. The Unit for MAE is seconds.
    }
    \label{tab:ckpt_pred}
    \resizebox{0.45\textwidth}{!}{
\begin{tabular}{l rrr}
\toprule
\textbf{Regression Model}  & \textbf{Input Feature}  & \textbf{K-fold MAE} & \textbf{Test MAE} \\ \midrule
\textbf{Univariate} & $S_{c}$                                                         & 0.345 ($\pm$ 0.099)  & 0.356         \\
\textbf{Multivariate} & $S_{d}$, $S_{m}$                                               & 0.291($\pm$ 0.139)   & 0.353         \\
\textbf{Multivariate, Two Components PCA} & $S_{d}$, $S_{m}$, $S_{i}$ & 0.286 ($\pm$ 0.142)  & 0.354         \\
\textbf{SVR RBF kernel} & $S_{c}$                                                          & 0.198 ($\pm$ 0.135)  & 0.245        \\ \bottomrule
\end{tabular}}
\end{table}

As shown in Table~\ref{tab:ckpt_pred}, the SVR model with RBF kernel yielded the best
MAEs for both k-fold cross validation and on the test dataset. The mean absolute
error percentage of the SVR model with RBF kernel on the test dataset is 5.38\%. The
other three models have up to 1.74X higher k-fold MAEs and around 1.45X higher 
test MAEs. All four models would have reasonable utility in
predicting total training time. 
For example, in the case of \emph{ResNet-32} that trains to $64K$ steps with $4K$
checkpoint interval, with linear regression model the actual and predicted checkpoint time are $3.83$ and
$3.96$ seconds---a difference of 3.4\%. 
Even though the prediction error is accumulative, it has
minimal impact on the final training time that is in the order of magnitude
of hours. 

Finally, practitioners might decide to choose a prediction model based on
factors other than the prediction accuracy, such as the time to retrain the
model.
For instance, if the practitioner is monitoring a running cluster and observes
variable performance, then the prediction model needs to be retrained with new
measurement data. It might be better to choose models that can be retrained
faster, e.g., multivariable models instead of SVR models, as the latter requires
hyperparameter tuning. 
\section{Characterizing Revocation Overhead}
\label{sub:revocation_overhead}

One of the key challenges of using transient servers for distributed training is
that they can be revoked at any time. 
Even the revocation of a single worker can lead to significant performance
degradation~\cite{2019icac:speedup}.
In this section, we characterize the revocation patterns of Google Cloud's
transient servers. In summary, we observed that cloud region, GPU type, and time-of-the-day are
important factors for understanding revocation patterns. 
Further, we found that immediately requesting a replacement worker after a
revocation is a valid strategy as the time to request transient GPU servers is
not impacted by revocations. 
Lastly, the workload of a transient server does not appear to impact its
likelihood of revocation. 

\subsection{Measurement Methodology}
\label{subsec:revocation:methodology}

\paragraph{\sysname Measurement Infrastructure.}
To measure the revocation of Google transient servers, we implemented a hook
function in TensorFlow in conjunction with startup and
shutdown scripts provided by Google Cloud. Each GPU worker in
the training cluster connected to the \sysname \emph{controller} running in the parameter server via RPC.
\emph{Transient-TensorFlow}, running on the GPU workers, monitored the
triggering of each script and forwarded the corresponding timestamped signals to
the \emph{controller}.

\paragraph{Measuring Transient Startup Time.}
\emph{Transient server startup time} is defined as the time between when
the cloud customer requested the transient server and when the transient server
became available in the training cluster.
For each transient server, we measured the time for three consecutive
stages~\cite{gce_life_cycle}. First, resources are allocated for the server during the
\emph{provisioning} stage. Second, after resource acquisition, the instance is
prepared for booting in the \emph{staging} phase. Third, once the server boots
up, it enters the \emph{running} stage. 
We used the Google Cloud API in conjunction with the startup script to
request servers and measured the duration for each stage by periodically
querying the cloud-returned state information.  
For each GPU-region combination, we requested 
transient and the equivalent on-demand servers for comparison.
To quantify \emph{availability-related}
startup overheads, we measured the time to start different
transient GPU servers after a predefined time window upon a revocation event through \sysname.


\begin{figure}[t]
    \centering
    \includegraphics[width=0.4\textwidth]{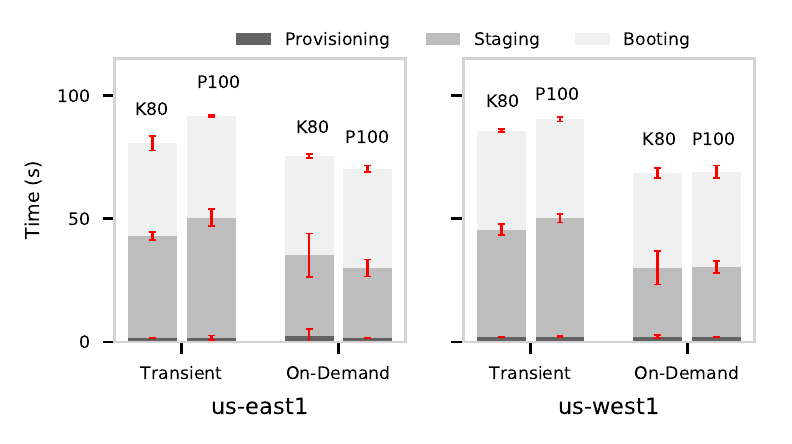}
    \caption{\textbf{Startup time breakdown of requesting new servers without observing revocation.} We requested \emph{K80} and \emph{P100} GPUs in \emph{us-east1} and \emph{us-west1} regions, with both transient and on-demand servers.}
    \label{fig:startup_norev}
\end{figure}

\begin{figure}[t]
    \centering
    \includegraphics[width=0.4\textwidth]{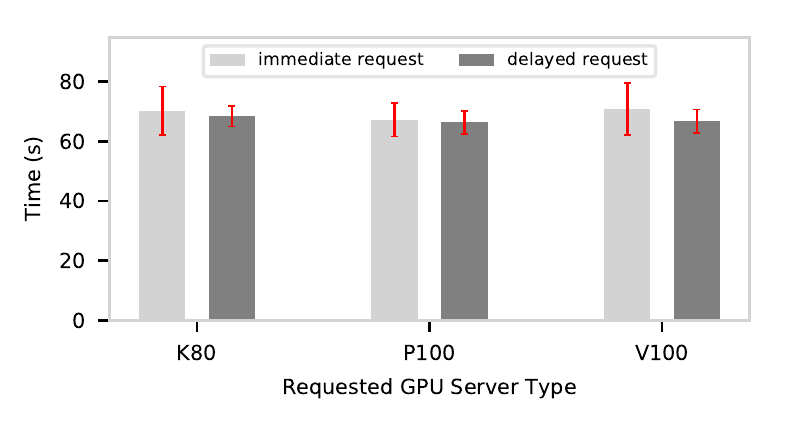}
    \caption{
    \textbf{Startup time comparison.} 
    We studied whether revocation events impact server startup time with
    immediate and delayed acquisition requests.
    }
    \label{fig:startup_rev}
\end{figure}


\begin{figure*}[t]
    \captionsetup[subfigure]{aboveskip=-1pt,belowskip=-1pt}
        \begin{subfigure}{0.3\textwidth}
        \centering
        \includegraphics[width=\textwidth]{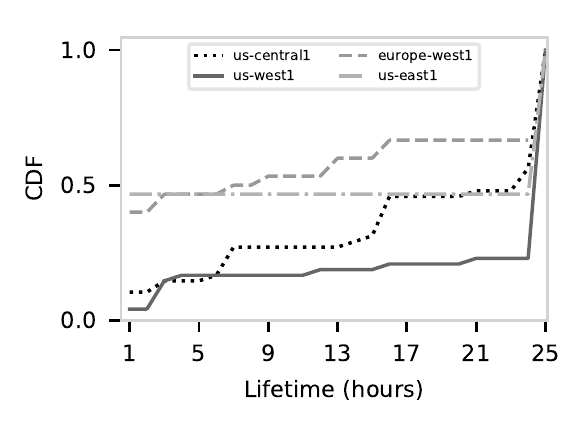}
            \caption{K80 GPU server.}
        \label{subfig:k80_life} 
        \end{subfigure}
    \hfill
        \begin{subfigure}{0.3\textwidth}
        \centering
        \includegraphics[width=\textwidth]{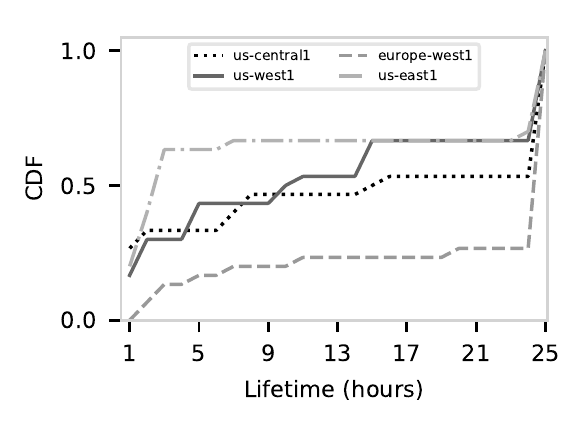}
            \caption{P100 GPU server.}
        \label{subfig:p100_life} 
        \end{subfigure}
    \hfill
        \begin{subfigure}{0.3\textwidth}
        \centering
        \includegraphics[width=\textwidth]{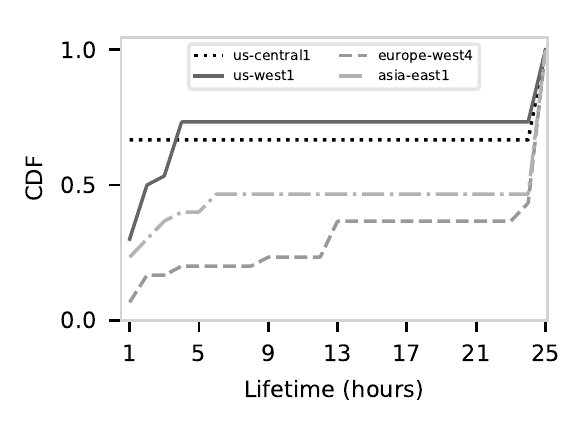}
            \caption{V100 GPU server.}
        \label{subfig:v100_life} 
        \end{subfigure}
        \caption{\textbf{Lifetime analysis for different regions and GPU
        servers.} We launched three types of transient GPU servers in six data centers throughout 12 non-consecutive days. We observed that
        up to 47.98\% lived up to, and even beyond, the specified
        maximum of 24 hours.
        }
        \label{fig:lifetime}
    \end{figure*}


\paragraph{Measuring Revocations.} 
We requested transient
GPU servers in batches. For each batch we requested the maximum number of
servers allowed for our account. We let these servers run for their maximum
lifetime of 24 hours and recorded any revocations that occurred prior to the
24-hour cutoff. 
We repeated this process for a total of twelve non-consecutive days. 
We divided the transient servers into two \emph{equally-sized groups}: the first group
contained idle servers and the second group consisted of servers that were
stressed in CPU, memory, and GPU resources. For stressing CPU and
memory resources, we used a popular benchmark~\cite{stress_ng} and for stressing
the GPU, we used built-in TensorFlow tasks that performed operations similar to
distributed training workloads. 
We repeated the above measurements for the three GPU types described previously
in six geo-graphically distributed regions---three US-based regions, two
Europe-based regions, and one Asian-based region---to study the impact of the
region and time-of-day on revocations.

\paragraph{Measuring Worker Replacement Overhead.}
\emph{Worker replacement overhead} denotes the time of
configuring the environment for distributed training after a worker replacement.
This includes starting the deep learning framework, joining the existing
training session, downloading the training dataset that the revoked server held,
and recomputing from the last checkpoint if needed. 
We measured the \emph{cold start} and \emph{warm start} worker replacement overhead with a cluster compromised of one \emph{K80} GPU worker and one parameter server. 
Cold start refers to the overhead when using a newly requested GPU server while warm start uses an existing GPU server. 

\paragraph{Measuring Recomputation Overhead.}
We trained \emph{ResNet-15} with 2-worker clusters and configured the checkpoint
interval to be $4$K steps. We manually revoked the 
chief worker at $1$K steps since the last checkpoint, and 
added a new worker to the training session at a specified interval.
In particular, \emph{recomputation overhead} denotes the time difference between adding a replacement worker with the chief's old IP address and adding a replacement worker with a new IP address.


\begin{table}[t]
\centering
    \sf
    \caption{
    \textbf{Transient GPU servers revocations by regions.}
    We observed that revocations vary based on GPU types and regions, e.g.,
    us-west1 region has the highest overall revocation percentage. Further,
    idle and stressed servers exhibit similar revocation rates. 
    }
    \label{tab:rev_reg}
    \resizebox{0.35\textwidth}{!}{
    \begin{tabular}{ l | r r r }
    \toprule
    \textbf{Regions}      & \textbf{K80}       & \textbf{P100}     & \textbf{V100}     \\ \midrule
    \textbf{us-east1}     & 30 (46.67\%)  & 30 (70\%) & N/A        \\ 
    \textbf{us-central1}  & 48 (56.25\%) & 30 (53.33\%)  & 30 (66.67\%) \\ 
    \textbf{us-west1}     & 48 (22.92\%)  & 30 (66.67\%) & 30 (73.33\%) \\ 
    \textbf{europe-west1} & 30 (66.67\%)  & 30 (26.67\%) & N/A        \\ 
    \textbf{europe-west4} & N/A         & N/A        & 30 (43\%) \\ 
    \textbf{asia-east1}   & N/A         & N/A        & 30 (47\%) \\ 
    \hline
    \textbf{total}        & 156 (46.15\%) & 120 (54.17\%) & 120 (57.5\%) \\ \bottomrule
    \end{tabular}}
    \end{table}
    

\subsection{Breaking Down Transient Startup Time}
\label{subsec:transient_provisioning}

Intuitively, transient startup time impacts transient distributed training
because it is the amount of time the training cluster has to run with
\emph{fewer} GPU workers after a revocation. In this subsection, we focus on quantifying the transient
startup time under different scenarios, such as immediately after server
revocations. Our understandings can help deep learning practitioners make
informed decisions about provisioning transient GPU servers.

Figure~\ref{fig:startup_norev} shows the average startup time of transient and
on-demand GPU servers in two cloud regions. 
Our \emph{first} observation is that it takes less than 100 seconds to startup
transient GPU servers. 
This short startup time makes it feasible for practitioners to quickly react to
a training slowdown by requesting and adding transient servers to
the ongoing training session.
\emph{Second,} it is on average 8.7\% slower to startup the more
powerful transient \emph{P100} GPU servers than \emph{K80} GPU servers, with the
staging time contributing most to the difference. 
The longer and more variable staging
time for transient \emph{K80} might be an indication of higher demand and
lower availability of \emph{K80} GPUs. 
\emph{Third,} compared to their
on-demand counterparts, transient startup time was only 11.14 seconds
slower on average for \emph{K80} and 21.38 seconds for \emph{P100} servers. 
Such slowdown is negligible for distributed training workloads which often last
hours if not days~\cite{coleman2017dawnbench,jeffdean}. 

\begin{figure*}[t]
    \captionsetup[subfigure]{aboveskip=-1pt,belowskip=-1pt}
        \begin{subfigure}{0.3\textwidth}
        \centering
        \includegraphics[width=\textwidth]{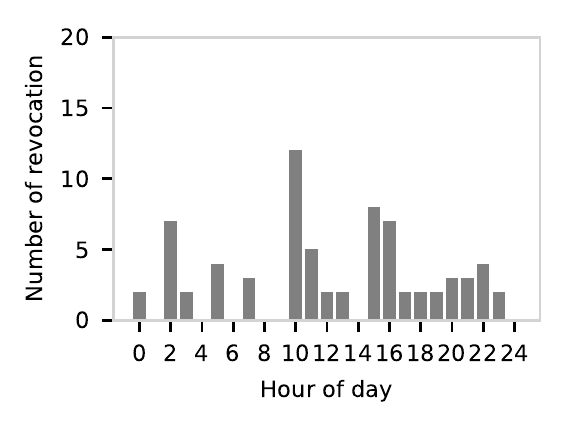}
            \caption{K80 GPU server.}
        \label{subfig:k80_day} 
        \end{subfigure}
    \hfill
        \begin{subfigure}{0.3\textwidth}
        \centering
        \includegraphics[width=\textwidth]{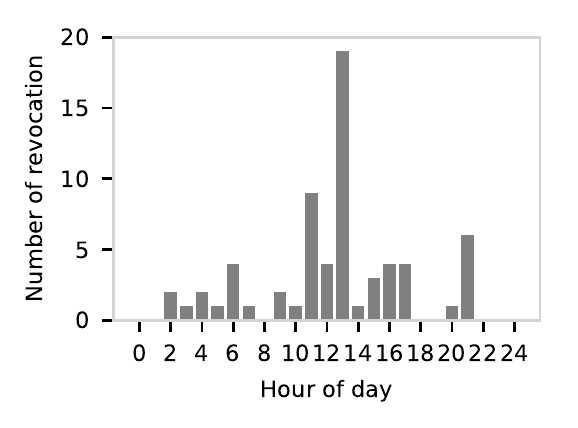}
            \caption{P100 GPU server.}
        \label{subfig:p100_day} 
        \end{subfigure}
    \hfill
        \begin{subfigure}{0.3\textwidth}
        \centering
        \includegraphics[width=\textwidth]{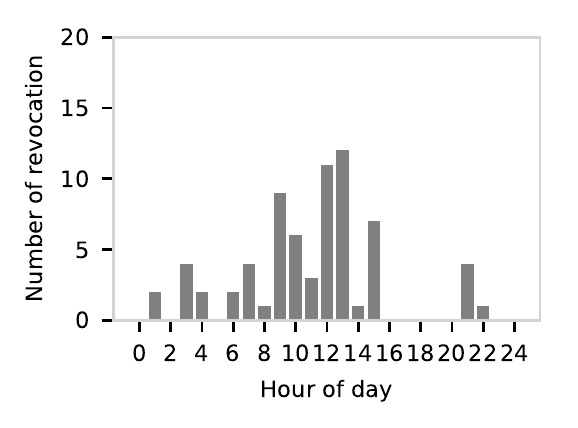}
            \caption{V100 GPU server.}
        \label{subfig:v100_day} 
        \end{subfigure}
        \caption{\textbf{Time-of-day impact on revocations.} We analyzed the
        revocation events for three types of GPU servers that were launched
        throughout twelve non-consecutive days. We found that revocations are
        both server and time-of-day dependent.}
        \label{fig:revocation_time_of_day}
    \end{figure*}
    

Figure~\ref{fig:startup_rev} shows the impact of recent revocations on transient
startup time. In particular, we studied \emph{immediate requests} and
\emph{delayed requests}. For the former, we
immediately requested a \emph{K80}, a \emph{P100}, and a \emph{V100} GPU server
after one of our \emph{K80} servers was revoked. Delayed requests are the same
as immediate requests except
that we waited for at least an hour before requesting. 

We observed little impact, up to 4 seconds in the case of \emph{V100} GPU
servers, of revocation events on transient startup time. 
These results are counter-intuitive as one of the potential reasons for
revocation is higher demand for a given resource~\cite{spotlight}.
These results suggest that deep learning practitioners do not need to consider
the revocation overhead associated with low availability.  
Further, the average startup time for immediate requests for both \emph{P100}
and \emph{V100} are within 3 seconds to that of \emph{K80}. This
suggests the possibility to request any GPU type as replacement for the revoked
server. 
The average startup time for immediate and delayed requests are within 4 seconds for all GPU types. 
However, for immediate requests, we observed a 4X higher coefficient of variance
(12\% compared to 3\%)---startup time is more variable immediately after a
server revocation.

\subsection{Understanding Transient Revocations}
\label{subsec:revocation_frequency}

Next, we looked at the different factors that impact the revocation frequency. 
Table~\ref{tab:rev_reg} summarizes the 206 revocations for 
\emph{396} transient GPU servers launched throughout twelve non-consecutive days, in six
different data centers.
Our \emph{first} observation is that the workload of
transient servers does not seem to impact the revocation frequency;
roughly half of all observed revocations were for unstressed servers, i.e., idle
servers.
Our \emph{second} observation is that different regions can lead to different revocation
frequencies. For example, \emph{europe-west1} has the lowest revocation frequency
for \emph{P100} while \emph{us-west1} region has the highest revocation
frequency for \emph{P100} and \emph{V100} GPU servers. As a simple
strategy, deep learning practitioners can avoid high revocation regions to
mitigate the impact on distributed training.
\emph{Third,} more expensive GPU servers, i.e.,
\emph{V100}, are more likely to be revoked compared to cheaper GPU
servers. 
This suggests the need to balance computation needs and revocations when
choosing GPU servers.

Figure~\ref{fig:lifetime} shows that different GPU servers in
different regions tend to have distinct lifetime characteristics. For example, 
more than 50\% of \emph{K80} servers from \emph{europe-west-1} were revoked in
the first two hours compared to less than 5\% from \emph{us-west-1}.
The mean time to revocation for \emph{K80} ranges from 10.6 hours to
 19.8 hours. This suggests the benefits of launching training clusters in regions such as \emph{us-central1} when using \emph{K80}. In addition, more
 powerful GPU servers tend to have a shorter mean time to revocation, e.g.,
 \emph{V100} servers in \emph{us-central1} had a mean time to revocation of 7.7 hours. 
 Combined, these observations also indicate the challenge of selecting the
 initial cluster configuration---a region that provides more stable \emph{K80} servers might have volatile \emph{V100} servers.

Figure~\ref{fig:revocation_time_of_day} illustrates the hour of the day when revocations occurred, represented in each region's local time.
Each GPU type exhibited different revocation patterns. 
For example, \emph{K80} servers had the highest number of revocations at 10AM,
perhaps caused by a surge of demand, while no revocations were observed 
for \emph{V100} servers between 4PM and 8PM.

Finally, our observations suggest an avenue for future work: 
investigating how strategically launching transient clusters at different times of day and different data center locations can help mitigate revocation impacts.

\subsection{Worker Replacement Overhead}
\label{subsec:restarting_overhead}

\begin{figure}[t]
    \centering
    \includegraphics[width=0.4\textwidth]{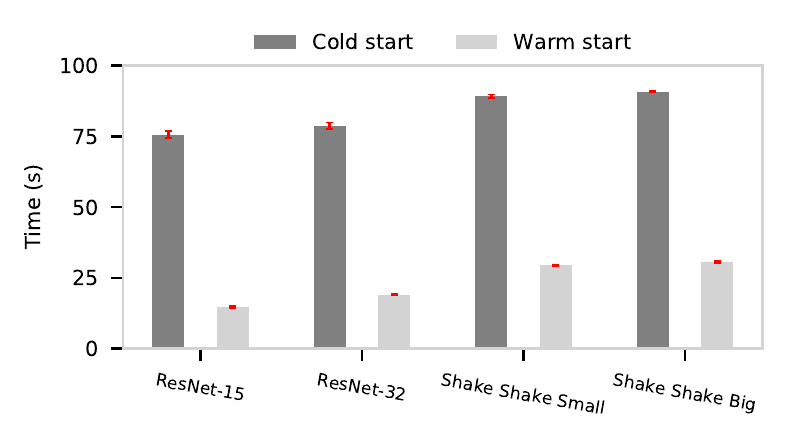}
    \caption{\textbf{Worker Replacement Overhead.} Cold start refers to the overhead when using a newly requested GPU server while warm start uses an existing GPU server.}
    \label{fig:tf_restart}
\end{figure}

Figure~\ref{fig:tf_restart} compares the cold start and warm start worker replacement time.
We make two key
observations. \emph{First,} requesting new workers after revocations (cold start) are much
more costly than scenarios where only restarting the training framework (warm start) is needed. 
For example, in the case of
\emph{ResNet-15}, it took about $75.6$ seconds compared to $14.8$ seconds. 
\emph{Second,} both the cold and warm start time increase with model size
and complexity. For instance, the worker replacement overhead for \emph{Shake Shake Big} was $15$ seconds longer than \emph{ResNet-15}, with most of the overhead coming from the training computation graph setup. 

We expect to observe similar overheads for \emph{P100} and
\emph{V100} clusters given that such overheads are not
GPU-dependent.

\subsection{TensorFlow-specific Recomputation Overhead}

In unmodified TensorFlow, we observed the following phenomenon: 
when the chief worker is revoked and a replacement worker is assigned the
chief's previous IP address, the cluster will recompute from the last
checkpoint. 
In other words, the cluster will discard any progress made since the last
checkpoint. 
By design, the IP address is bound to the role of chief. Therefore, the 
replacement worker effectively becomes the new chief. 
As the chief worker is responsible for saving the checkpoint, the recomputation
overhead can be high. 
Note, \sysname's \emph{transient-tensorflow} avoids such overhead;
consequently, we do not consider this overhead in modeling distributed transient
training. 

\begin{figure}[t]
    \centering
    \includegraphics[width=0.4\textwidth]{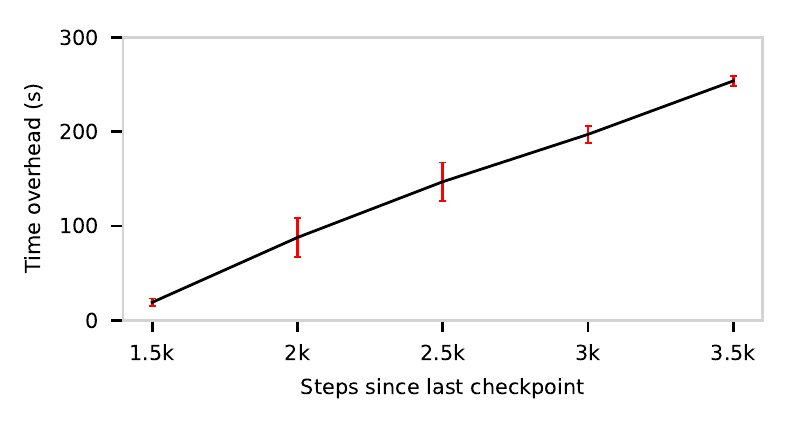}
    \caption{\textbf{TensorFlow-specific recomputation overhead.} 
    We plotted the time savings of assigning a new IP address
    to the replacement worker, 
    compared to reusing the old chief worker's IP address. 
    }
    \label{fig:restart}
\end{figure}

Figure~\ref{fig:restart} shows the recomputation overhead of
training \emph{ResNet-15} using a two-\emph{K80} GPU cluster.
We configured the checkpoint interval to be $4K$ and manually
revoked the chief worker $1K$ steps after the last checkpoint.
We evaluated the impact of the \emph{replacement timing}, i.e., when the
replacement worker is added and starts training.
For each replacement timing, we measured the total time to reach the next designated checkpoint, with and
without reusing the chief worker's IP and calculated the time difference (i.e.,
recomputation overhead). 
When using \sysname, an existing worker in the training session will be assigned
the responsibility of checkpoint and therefore the recomputation overhead is bounded by the checkpoint interval. 
In Figure~\ref{fig:restart}, such overhead is up to $224$ seconds with a $4K$
steps checkpoint interval. 


\section{Use cases of performance modeling}
\label{sec:use_cases}

Finally we discuss how practitioners might leverage the findings and
insights of our work for \1 predicting cluster training speed and \2
detecting training bottlenecks. 
These use cases represent promising extensions to the measurement study
presented in this work. Below we describe our preliminary evaluations but
leave a comprehensive analysis for future work.

\subsection{Heterogeneous Training Prediction}
\label{sec:usecase:heterogeneous}

To predict the speed of \emph{heterogeneous clusters}, i.e., clusters that consist of
different types of GPU servers, we can leverage the GPU
worker and parameter server performance models described in Section~\ref{sec:training_speed}. These models can be built offline using 
historical measurement data and retrained with continuous monitored data. 

In Section~\ref{sec:training_speed}, we observed that individual server training
speed can be predicted using CNN model complexity and the computational capacity
of the server's GPU. 
Further, we observed that adding GPU servers of different types to
an asynchronous training session will not impact existing GPU workers' training
speed. Therefore, we can predict the cluster
training speed as $sp=\sum_{i}^n sp_i$ for a cluster of $n$ GPU servers, where
$sp_i$ denotes the training speed of GPU server $i$.
The predicted training time $T$ for $N_{w}$ amount of training work,
measured in \emph{number of training steps}, is then: 

\setlength{\abovedisplayskip}{0pt} 
\setlength{\belowdisplayskip}{0pt}
\begin{align} 
    T & = \frac{N_{w}}{sp} + \big\lceil\frac{N_w}{I_c}\big\rceil \times T_{c} +
    N_r \times (T_{p} + T_{s}), \label{eq:usecase:prediction} \\ 
    N_r & = \sum\limits_{i}^n Pr(R_i), \label{eq:usecase:revocation}
\end{align}

where $I_c$, $T_c$, $T_p$, and $T_s$ denote the checkpoint interval (number of
steps), checkpoint time, time to provision a new GPU server, and worker
replacement time, respectively. We assume that $N_w$ and $I_c$ are user-specified
values, $sp$ and $T_c$ are predicted for CNN models given their FLOPs, and $T_p$
and $T_s$ are running averages based on historical measurements. 

The expected number of revocations $N_{r}$ is
calculated as the sum of the probabilities $Pr(R_i)$ that each worker $i$
will be revoked during the training. We obtain these probabilities by querying
the empirical CDFs, e.g, Figure~\ref{fig:lifetime}.
For simplicity, we do not consider the impact of newly added transient servers
on the number of expected revocations.
However, we have additional empirical data for supporting other more
complicated modeling scenarios. 

When using Equations~\eqref{eq:usecase:prediction}
and~\eqref{eq:usecase:revocation}, we observed a 0.8\% prediction error for
\emph{ResNet-32} with $N_w=64K$ and $I_c=4K$ using our measurements. 
In summary, this work and \sysname provides the
foundation for understanding and modeling transient distributed training performance.

\subsection{Detecting Training Bottlenecks}
\label{sec:uescase:bottleneck}

\begin{figure}[t]
        \begin{subfigure}{0.23\textwidth}
        \centering
        \includegraphics[width=\textwidth]{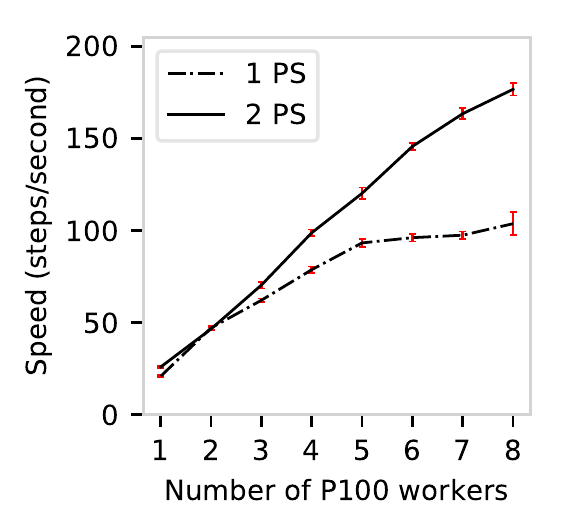}
            \caption{ResNet-15.}
        \label{subfig:ps_bot_2:resnet15} 
        \end{subfigure}
    \hfill
        \begin{subfigure}{0.23\textwidth}
        \centering
        \includegraphics[width=\textwidth]{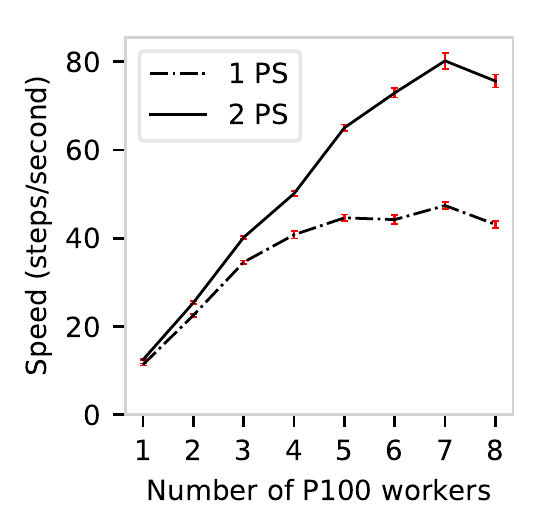}
            \caption{ResNet-32.}
        \label{subfig:ps_bot_2:resnet32} 
        \end{subfigure}
        \caption{
            \textbf{Parameter-server based bottleneck detection and mitigation.}
            We observed plateaued training speed for larger clusters with one
            parameter server. By adding a second parameter server, the training
            speed was improved by up to 70.6\%. Similar trend was observed for
            \emph{Shake Shake} models as well.
            }
        \label{fig:ps_bot_2}
    \end{figure}

Troubleshooting distributed training performance is challenging as
bottlenecks can be caused by a plethora of factors such as network variations
between parameter servers and GPU workers, and cloud server performance
fluctuations. 
We illustrate detecting one such bottleneck caused by overloaded parameter servers. 
However, we believe that our method and \sysname are extendable to detect and
resolve other bottlenecks.

Figure~\ref{fig:ps_bot_2} compares the training speed for clusters with one
parameter server and ones with two parameter servers. When training
\emph{ResNet} models using one parameter server, we observed that larger
clusters, e.g., with six \emph{P100} servers, do not yield reasonable speedup
compared to smaller clusters. 
Although sublinear scalability in distributed training is not
a myth~\cite{sergeev2018horovod, dl_perf2}, \sysname allows one to detect
\emph{when} such bottlenecks arise during training. 
For example, if the predicted theoretical training speed (as described in
Section~\ref{sec:usecase:heterogeneous}) and the measured one differ by 
a configurable threshold, \sysname will flag the bottleneck. Currently, we
use a warmup period of $30$ seconds and a threshold of 6.7\% based on
empirical observation. Similar approaches can be used to detect slower GPU
workers as well. 

One potential way to resolve this parameter-server-based bottleneck is to
increase the number of parameter servers to two. 
This improved the training speed of all clusters by up to 70.6\%. However,
currently deep learning frameworks such as TensorFlow do not support
dynamically adding parameter servers while training
is ongoing---one has to restart the training session which incurs an overhead
of about $10$ seconds. We leave overhead-aware bottleneck mitigation as future work. 

\section{Related work}
\label{sec:related}

\paragraph{Distributed Training.}
Cloud computing has become the \emph{de facto} platform for hosting a plethora
of modern applications, deep learning as an emerging workload is no exception~\cite{strom2015scalable}. Popular deep learning
frameworks~\cite{caffe2,tensorflow,cntk,mxnet} provide distributed
SGD-based algorithms~\cite{sgd1,stale4} to train increasingly bigger models on larger datasets. 
Existing works towards understanding distributed training workloads can be
broadly categorized into performance
modeling~\cite{dl_perf1,qi:iclr17:paleo,shi2018performance} and empirical
studies~\cite{zou2017distributed,coleman2017dawnbench,shi2018performance,2019icac:speedup,Jeon:atc2019:analysis}. 
In contrast to prior model-driven performance modeling
studies~\cite{lin2018model, zheng2019cynthia, qi:iclr17:paleo}, where a static
end-to-end training time prediction is the main focus, our work leverages
data-driven modeling that is powered by a
large-scale empirical measurement in a popular cloud platform. 
The insights provided from both theoretical and empirical characterizations of
distributed training lead to numerous system-level optimizations. For
example, prior
work~\cite{jiang2017heterogeneity,zhang:atc17:poseidon,Luo:socc2018:parameterhub,Xie:socc2018:Orpheus}
designed heterogeneity-aware distributed training
systems for handling shifted bottlenecks or
identifying remaining training workload.
Our work adds unique knowledge of distributed training with transient servers
and framework modifications for transient-aware training, which can be valuable
for resource managers. 

\paragraph{Optimization for Transient Servers.} Researchers have
proposed various system-level techniques such as dynamic
checkpointing~\cite{spotcheck,flint,spoton}, to
explore the economic benefit brought by cloud transient servers.
Additionally, prior work also accounted for application-specific requirements,
such as interactivity, when designing transient-aware
mechanisms~\cite{spotcheck,tributary}. As a promising and cheap way to provide
good parallelisms, transient servers have garnered a lot of interests for big
data big data analytics~\cite{See_spotrun,flint,spoton,ambati2019optimizing}, 
memory-intensive applications~\cite{spot_burstable}, cluster resource
managers~\cite{portfolio-driven,proteus}, and most recently deep
learning~\cite{2019icac:speedup}.
Our work provides a new perspective with a focus on  characterizing and modeling
distributed training on transient servers. 

\section{Summary}
\label{sec:conclusion}

We explored the characteristics of and key factors impacting \emph{distributed
training on transient servers}. 
We chose three commonly-used GPUs from six data center locations for measuring
and modeling the performance of twenty CNN models. We found that simple
regression-based models have adequate prediction accuracy, even for
heterogeneous clusters and when training CNN models with diverse
characteristics. Additionally, we demonstrated that the overhead of commonly
used fault-tolerance mechanisms (i.e., model checkpointing) can be predicted
with high accuracy and the associated impact can be directly added to the
predicted training time.
Lastly, we explored potential use cases of our performance modeling including 
detecting and mitigating performance bottlenecks.
We envision that our study, together with our
open-source data, lays the the foundation for future research in optimizing transient distributed training.

\section*{Acknowledgment} We would like to first thank all anonymous reviewers
for their insightful comments. 
This work is supported in part by 
National Science Foundation grants \#1755659 and \#1815619,
and Google Cloud Platform Research credits.

\balance
\footnotesize{
\bibliographystyle{IEEETran}
\bibliography{./spottrain}
}

\end{document}